\documentclass[twocolumn]{aastex631}
\usepackage{color}
\usepackage[titletoc]{appendix}
\usepackage{amsmath}
\usepackage{amssymb}
\usepackage{mathtools}
\usepackage{upgreek}
\usepackage{float}
\usepackage{comment}
\usepackage{enumitem}
\usepackage{natbib}
\usepackage{graphicx}
\usepackage{bm}
\usepackage{totcount}
\usepackage{multirow}

\newtotcounter{citnum} 
\def\oldbibitem{} \let\oldbibitem=\bibitem
\def\bibitem{\stepcounter{citnum}\oldbibitem}

\shortauthors{Sobski \& Millholland}

\begin{document} 

\title{Can Cold Jupiters Sculpt the Edge-of-the-Multis?}

\author{Nicole Sobski}
\affiliation{Department of Computer Science, Wellesley College, Wellesley, MA 02481, USA}
\affiliation{Kavli Institute for Astrophysics and Space Research, Massachusetts Institute of Technology, Cambridge, MA 02139, USA}

\author[0000-0003-3130-2282]{Sarah C. Millholland}
\affiliation{Kavli Institute for Astrophysics and Space Research, Massachusetts Institute of Technology, Cambridge, MA 02139, USA}
\affiliation{Department of Physics, Massachusetts Institute of Technology, Cambridge, MA 02139, USA}
\email{sarah.millholland@mit.edu}

\begin{abstract}
Compact systems of multiple close-in super-Earths/sub-Neptunes (``compact multis'') are a ubiquitous outcome of planet formation. It was recently discovered that the outer edges of compact multis are located at smaller orbital periods than expected from geometric and detection biases alone, suggesting some truncation or transition in the outer architectures. Here we test whether this ``edge-of-the-multis'' might be explained in any part by distant giant planets in the outer regions ($\gtrsim 1$ AU) of the systems. We investigate the dynamical stability of observed compact multis in the presence of hypothetical giant ($\gtrsim 0.5 \ M_{\mathrm{Jup}}$) perturbing planets. We identify what parameters would be required for hypothetical perturbing planets if they were responsible for dynamically sculpting the outer edges of compact multis. ``Edge-sculpting'' perturbers are generally in the range $P\sim100-500$ days for the average compact multi, with most between $P\sim200-300$ days. Given the relatively close separation, we explore the detectability of the hypothetical edge-sculpting perturbing planets, finding that they would be readily detectable in transit and radial velocity data. We compare to observational constraints and find it unlikely that dynamical sculpting from distant giant planets contributes significantly to the edge-of-the-multis. However, this conclusion could be strengthened in future work by a more thorough analysis of the detection yields of the perturbing planets.

\end{abstract}

\section{Introduction}
\label{sec: Introduction}

Short-period super-Earths/sub-Neptunes are the most prevalent class of observed extrasolar planets, orbiting around roughly half of Sun-like stars \citep[e.g.][]{2013PNAS..11019273P, 2018ApJ...860..101Z, 2019MNRAS.490.4575H}. First discovered by early Doppler surveys \citep[e.g.][]{2011arXiv1109.2497M} and then, to a greater degree, by NASA's Kepler Mission \citep{2010Sci...327..977B}, short-period planets are often found in tightly-spaced configurations, with multiple planets all having orbital periods in the range from days to months \citep{2011ApJS..197....8L, 2014ApJ...784...44L, 2014ApJ...784...45R, 2014ApJ...790..146F}. These so-called ``compact multiple-planet systems'' (or ``compact multis'') exhibit a remarkable degree of structure and regularity in both their orbital and physical properties. Their orbits are nearly coplanar and circular \citep[e.g.][]{2012ApJ...761...92F, 2014ApJ...790..146F, 2015ApJ...808..126V,2016PNAS..11311431X}. Moreover, their period ratios, radii, and masses exhibit a statistical tendency towards uniformity within a given system \citep{2018AJ....155...48W, 2017ApJ...849L..33M}. This set of patterns is known collectively as the ``peas-in-a-pod patterns'' or ``intra-system uniformity''  (for a review, see \citealt{2022arXiv220310076W}).

Despite the abundance of information the Kepler Mission delivered about the architectures of close-in super-Earths/sub-Neptunes, it was mostly limited to the inner regions ($a \lesssim$ 1 AU) of the systems. However, various observational and theoretical efforts are gradually revealing more about their outer architectures. For instance, using radial velocity (RV) observations to search for long-period companions,  \cite{2018AJ....156...92Z} and \cite{2019AJ....157...52B} showed evidence that distant giant planets (or ``cold Jupiters'', with $a \gtrsim 1$ AU, $M_p \gtrsim 0.5 \ M_{\mathrm{Jup}}$) are over-represented in systems with inner super-Earths. They found that $\sim30-40\%$ of systems with inner super-Earths are found to have distant giant planets, relative to the $\sim10\%$ distant giant occurrence for stars irrespective of small planet presence. The super-Earth/distant giant correlation was also studied by \cite{2022ApJS..262....1R} in their analysis of the California Legacy Survey \citep{2021ApJS..255....8R}; they found tentative agreement with the earlier studies, showing that stars with inner small planets may have an enhanced occurrence of outer giants with $1.7\sigma$ significance. 

A contradictory result was recently presented by \cite{2023arXiv230405773B}, who studied RVs of 38 Kepler and K2 small-planet systems collected over nearly a decade with the HARPS-N spectrograph, as well as publicly available measurements collected with other facilities. They detected five cold Jupiters in three systems and derived an occurrence rate of $9.3^{+7.7}_{-2.9}\%$ for planets with $0.3 - 13 \ M_{\mathrm{Jup}}$ and $1-10$ AU in systems with inner small planets. \cite{2023arXiv230405773B} found no evidence of the overabundance of distant giant planets in super-Earth systems claimed by \cite{2018AJ....156...92Z} and \cite{2019AJ....157...52B}. However, \cite{2023arXiv230616691Z} recently suggested that the discrepancy between the \cite{2023arXiv230405773B} result and previous works can be fully resolved by accounting for the metallicity dimension of the super Earth–cold Jupiter relations. Further constraints may come from the Kepler Giant Planet Survey (KGPS, \citealt{2023arXiv230400071W}), a decade-long survey of 63 Kepler systems using HIRES at the Keck Observatory designed to search for long-period planets in Kepler systems. \cite{2023arXiv230400071W} presented RV-detected companions to 20 stars in their sample, with occurrence rate analyses forthcoming.  It is clear that the relationship between inner super-Earths and outer giants is still debated but gradually coming into better focus.

Distant giant planets in general often have moderate to high eccentricities, $e \sim 0.1-0.8$ \citep[e.g.][]{2006MNRAS.369..249J, 2007ARA&A..45..397U, 2009ApJ...693.1084W, 2013ApJ...767L..24D}, and their orbits might be significantly mutually-inclined with respect to the inner systems \citep[e.g][]{2017MNRAS.464.1709G}. In some cases, an eccentric and/or inclined distant perturber can exert a stronger gravitational influence on the inner planets than their mutual gravitational coupling, and the inner system can gain significant eccentricities and inclinations, potentially driving dynamical instabilities \citep[e.g.][]{2017AJ....153...42L, 2017MNRAS.467.1531H, 2018MNRAS.478..197P, 2019MNRAS.482.4146D, 2020AJ....160..105S}. The dynamics of the inner system can thus be tightly coupled to that of the outer system. 

Given this dynamical interaction, if a system has both small inner planets and a distant giant planet, it must experience an architectural transition point in its overall layout. The transition point could be generated by the perturbative influence of the distant giant planet. Specifically, each system with an eccentric and possibly inclined distant giant planet must possess a ``zone of instability'', where any planets existing in that region would experience eccentricity/inclination excitation and destabilization. It is possible that the impacts of this are already detectable. 

One potential link is the result of \cite{2022AJ....164...72M}, who presented statistical evidence for an outer truncation or transition (named the ``edge-of-the-multis'') in the observed high-multiplicity Kepler systems. They explored the existence of hypothetical additional super-Earths/sub-Neptunes orbiting beyond the outermost planets in Kepler compact multis, under the assumption that the ``peas-in-a-pod'' patterns continue to larger separations than observed. They then estimated the transit and detection probabilities of these hypothetical planets and found that additional exterior planets should have been detected in $\gtrsim35\%$ of Kepler compact multis. Accordingly, to $\gtrsim7\sigma$ confidence, the Kepler compact multis truncate at smaller orbital periods than expected from geometric and detection biases alone. These results can be explained if compact multis experience an architectural truncation (i.e. an occurrence rate decrease) or a transition (i.e. a breakdown in the ``peas-in-a-pod'' patterns) at $\sim100-300$ days.

Although various theories predict an ``edge-of-the-multis'' consistent with \cite{2022AJ....164...72M}'s observation \citep[e.g.][]{2014ApJ...780...53C, 2022ApJ...937...53Z, 2023NatAs...7..330B}, it is still unknown what physical mechanism(s) are most responsible. In this paper, we explore the possibility of a link between the outer edges of compact multis and distant giant planets. That is, we test the hypothesis that dynamical perturbations from distant giant planets are strong enough to sculpt the outer edges. We explore this with the following steps: (1) We first consider the observed properties of compact multis and identify where hypothetical distant giant planets with various properties could exist without causing dynamical instabilities. (2) Next, we determine the required properties of the distant giant planets if they were sculpting the outer edges. (3) Finally, we compare the properties of these ``edge-sculpting'' distant giants to observational constraints. It is important to note that distant giants cannot be the \textit{only} explanation of the edge-of-the-multis, since there are not enough of them in compact multi systems (even if they were as prevalent as suggested by \citealt{2018AJ....156...92Z} and \citealt{2019AJ....157...52B}). Our aim here is to address whether they could play any significant role in the edge-of-the-multis.

This paper is organized as follows. We begin by describing our planet sample and our methodology for assessing dynamical stability of compact multis in the presence of distant giant planets (Section \ref{sec: methods}). We then present the results of these calculations, both of individual systems and the whole population (Section \ref{sec: results}). This allows us to determine the properties of perturbing planets that would be capable of sculpting the outer edges of the compact multis. We explore the detectability of these perturbing planets and the resulting implications for the edge-of-the-multis (Section \ref{sec: discussion}). We discuss alternative theories for the edge-of-the-multis and conclude in Section \ref{sec: conclusion}. 

\section{Methods}
\label{sec: methods}

\subsection{Planet sample}
\label{sec: planet sample}

\begin{figure}
    \centering
    \includegraphics[width=1.05\columnwidth]{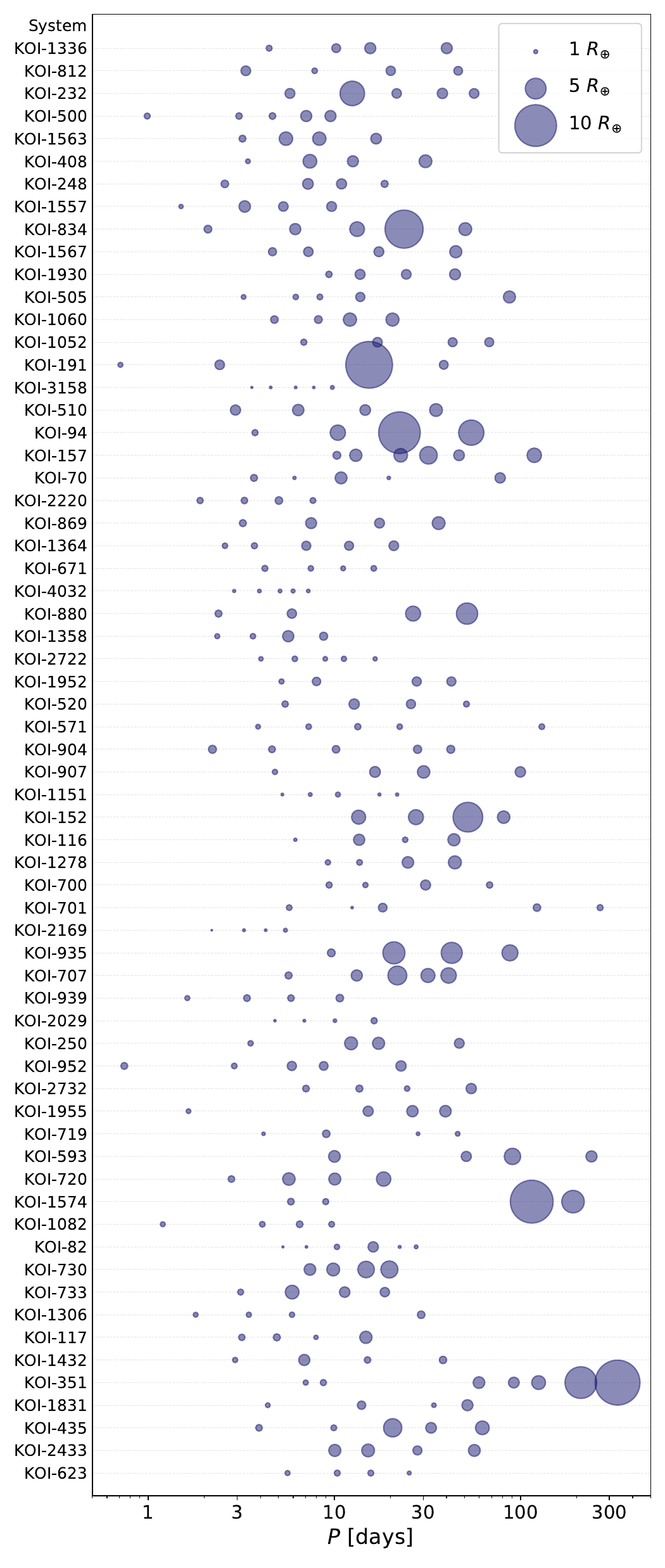}
    \caption{\textbf{Planet sample.} We display the architectures of the 64 Kepler compact systems with four or more transiting planets described in Section \ref{sec: planet sample}. The dot size is proportional to the planet size, as shown in the legend at top.}
    \label{fig: planet sample}
\end{figure}

\begin{table*}[t!]
\caption{\textbf{Parameter space scope.} Parameters used in the dynamical stability calculations and their sampling scheme.}
\centering
\begin{tabular}{l l l}
\hline
\hline
Parameter & Inner planets & Outer perturber \\ 
\hline
Mass & $M_p \sim \mathrm{Normal}(\overline{M}_p, \sigma_{M_p})$ & $M_{p,p} \sim \mathrm{Uniform}[0.5 \ M_{\mathrm{Jup}}, 5 \ M_{\mathrm{Jup}}]$ \\ 
Semi-major axis & $a = $ observed & $a_p \sim \mathrm{Uniform}[a(P_p = 1.3 P_{\mathrm{out}}), a(P_p=1000 \ \mathrm{days})]$ \\ 
Eccentricity & $e \sim \mathrm{Rayleigh}(0.04)$ & $e_p \sim \mathrm{Uniform}[0, 0.7]$ \\
Inclination & $i \sim \mathrm{Rayleigh}(1^{\circ})$ & $i_p \sim \mathrm{Rayleigh}(10^{\circ})$ (Case 1); $i_p \sim \mathrm{Uniform}[0^{\circ}, 40^{\circ}]$ (Case 2) \\ 
Argument of periapse & $\omega \sim \mathrm{Uniform}[0^{\circ},360^{\circ}]$ & $\omega_p \sim \mathrm{Uniform}[0^{\circ},360^{\circ}]$ \\ 
Longitude of ascending node & $\Omega \sim \mathrm{Uniform}[0^{\circ},360^{\circ}]$ & $\Omega_p \sim \mathrm{Uniform}[0^{\circ},360^{\circ}]$ \\ 
Mean anomaly & $M \sim \mathrm{Uniform}[0^{\circ},360^{\circ}]$ & $M_p \sim \mathrm{Uniform}[0^{\circ},360^{\circ}]$ \\

\label{tab: parameter space scope}
\end{tabular}
\end{table*}

We begin by defining our sample of Kepler compact multiple-planet systems, which is identical to the sample studied in \cite{2022AJ....164...72M}. We consider systems with four or more observed transiting planets only, since it is a well-defined high multiplicity sample that is most suitable to investigations of the outer edges. As our starting point, we use the Kepler DR25 KOI catalog \citep{2018ApJS..235...38T, koidr25} and consider all planets with ``confirmed'' and ``candidate'' dispositions. Where possible, we replace the stellar parameters and planet radii in the DR25 catalog with parameters from the Gaia-Kepler Stellar Properties Catalog \citep{2020AJ....160..108B, 2020AJ....159..280B}. We apply two quality cuts. First, we consider only planets smaller than $16 \ R_{\oplus}$ with fractional radius uncertainties less than 100\%. Second, we discard targets for which \cite{2017AJ....153...71F} found a companion star that contributed more than 5\% of the light in the photometric aperture. For the purposes of this study, we consider only systems with four or more observed transiting planets, which represent a high fidelity sample of compact multis. We are left with 279 planets in 64 Kepler systems with four or more observed transiting planets. The planet sample is displayed in Figure \ref{fig: planet sample}.

Most planets in these systems do not have measured masses due to observational limitations. (For instance, the star is too faint for Doppler follow-up and/or there are no detectable transit timing variations.) However, in cases where they are available (60 of 279), we utilize observational mass estimates from radial velocity or transit timing variation analyses, which we obtain from the \cite{PSCompPars}. For the remaining planets, we obtain approximate mass estimates from the \texttt{Forecaster} probabilistic mass-radius prediction model by \cite{2017ApJ...834...17C}. \texttt{Forecaster} is an open-source package that predicts a missing planetary mass or radius based on a mass-radius relationship established from a sample of 316 objects of various sizes spanning small planets to late-type stars. Altogether, our approach provides us with a heterogeneous collection of mass estimates, which may affect our results at the detailed level. However, this approach is sufficient for our purposes insofar as we are obtaining an average constraint across the compact multis as a population. 

\subsection{Dynamical stability calculations without perturbers}
\label{sec: dynamical stability calculations w/o perturbers}
We first assess the dynamical stability of the compact multi-planet systems before introducing additional perturbing planets. These estimates will serve as our basis of comparison to the later calculations. We perform our dynamical stability analyses using SPOCK (Stability of Planetary Orbital Configuration Klassifier; \citealt{2020PNAS..11718194T}), a machine learning model that predicts dynamical stability over long timescales ($\sim10^9$ orbits) by first running a short integration ($10^4$ orbits), assessing summary statistics, and classifying the stability using the XGBoost machine learning algorithm. SPOCK provides dynamical stability estimates at a rate that is five orders of magnitude faster than direct $N$-body integrations. Here, we will utilize the SPOCK-derived probability of stability, which we denote $p_{\mathrm{stab}}$.

To estimate $p_{\mathrm{stab}}$, SPOCK requires the masses and orbital elements as inputs. Our planet sample only has measured stellar masses, planetary masses (most of which are estimated using \texttt{Forecaster}, as described previously), and periods, so we must randomly sample the remaining parameters. We take the orbital eccentricities and inclinations to be Rayleigh-distributed with scale parameters consistent with current observational constraints \citep{2016PNAS..11311431X, 2019AJ....157...61V, 2019AJ....157..198M}. We take the arguments of periastron, longitudes of the ascending node, and mean anomalies to be uniformly distributed from $0^{\circ}$ to $360^{\circ}$. Because the planet masses are subject to uncertainties (especially in the case where the masses are predicted from the \texttt{Forecaster} model), we include random sampling of the masses. We take the masses to be normally distributed around the mean observed or estimated value, $\overline{M}_p$, with standard deviations $\sigma_{M_p}$ equal to the average of the observational uncertainties $\sigma_{M_p} = (\sigma_{M_p,\mathrm{high}}+ \sigma_{M_p,\mathrm{low}})/2$ in the cases where the mass estimates are from RVs or TTVs, or equal to $0.3\overline{M}_p$ in the cases where the mass estimates are from \texttt{Forecaster}. The details of the parameter sampling are summarized in Table \ref{tab: parameter space scope}. This sampling is performed for each planet in each system in our sample. New random variables are drawn for each iteration of dynamical stability calculations, as will be described in further detail in the next section.


For each system, we estimate $p_{\mathrm{stab}}$ 100 times, with each iteration having a different set of random variables. We then find the mean probability of stability, $\overline{p}_{\mathrm{stab}}$ and consider that to be representative of the system. Although a majority of the compact multi-planet systems are classified as stable, there are 20 systems with $\overline{p}_{\mathrm{stab}} < 0.5$, with 6 of these systems demonstrating $\overline{p}_{\mathrm{stab}} < 0.1$. These systems are preferentially near mean-motion resonances.\footnote{Considering the subset of unstable systems, we found that 45\% of adjacent planet pairs are near a first-order mean-motion resonance, compared to 15\% of planet pairs within the subset of stable systems. Moreover, 95\% of the unstable systems contain at least one near-resonant pair compared to 40\% of the stable systems.} This is unsurprising given that we are randomly sampling the mean anomalies, and planets can become destabilized near resonances if they not placed in specific parts of their orbits. Additional unstable systems that are not near-resonance may have a heightened sensitivity to mass variability (e.g. the real masses may be significantly smaller than the estimated masses), and the combination of this and the randomization of the other orbital elements could yield instability. We disregard these systems in our subsequent analysis, but we believe our analysis is still robust as a population average over many systems. We are more interested in the influence of distant perturbing planets on the ensemble of compact multis than on individual systems.

\subsection{Dynamical stability calculations with perturbers}
\label{sec: distant perturbers}
Having assessed the dynamical stability of the inner systems in isolation, we are ready to test the limits of additional outer planets in each stable system. We again opt to use SPOCK for our stability calculations. However, an important caveat is that the system architectures with both inner multis and distant giant planets are somewhat outside of SPOCK's training models, which only considered masses up to $0.1 \ M_{\mathrm{Jup}}$ and inclinations $\lesssim 10^{\circ}$ \citep{2020PNAS..11718194T}. Although SPOCK has been shown to generalize reasonably well, it is still important to check its validity in this context. We performed multiple checks using other numerical and analytical stability estimates, which we describe in more detail in Appendix \ref{sec: testing SPOCK}.

To assess the dynamical impact of the outer perturbing planet, we randomly sample its physical and orbital properties within reasonable ranges. The outer perturber is taken to be a massive gas giant planet on an eccentric and moderately inclined orbit, in line with general observed properties of cold Jupiters \citep[e.g.][]{2013ApJ...767L..24D}. We sample its mass as $M_{p,p} \sim \mathrm{Uniform}[0.5 \ M_{\mathrm{Jup}}, 5 \ M_{\mathrm{Jup}}]$. We consider periods, $P_p$, between $1.3 P_{\mathrm{out}}$ (where $P_{\mathrm{out}}$ is the outermost observed planet in the inner system) and $1000$ days. These limits were chosen because $1.3 P_{\mathrm{out}}$ is approximately the smallest possible stable period, and $1000$ days is sufficiently distant to detect the transition between stable and unstable orbits. The $1000$ day maximum period also extends beyond the maximum detectable window via transit detection. We sample uniformly in values of the perturber's semi-major axis, $a_p$, according to this period range. 

The outer perturber's eccentricity is sampled as ${e_p \sim \mathrm{Uniform}[0,0.7]}$, which encompasses the range of eccentricities of most observed distant giant planets \citep[e.g.][]{2021ApJS..255....8R}. As for the perturber's inclination, we consider two different cases. In Case 1, we consider the perturber to be nearly coplanar with the inner system and sample $i_p \sim \mathrm{Rayleigh}(10^{\circ})$. This is motivated by \cite{2020AJ....159...38M}, who suggested that distant giant planets orbiting compact multis tend to be nearly coplanar with the inner systems. In Case 2, we consider a wider range of inclinations for the perturber, $i_p \sim \mathrm{Uniform}[0^{\circ},40^{\circ}]$. This range is consistent with the expectations of energy equipartition between eccentricities and inclinations. The remaining orbital elements (argument of periastron, longitude of the ascending node, and mean anomaly) are sampled uniformly from $0^{\circ}$ to $360^{\circ}$.

\begin{figure*}
    \centering
    \includegraphics[width=0.8\textwidth]{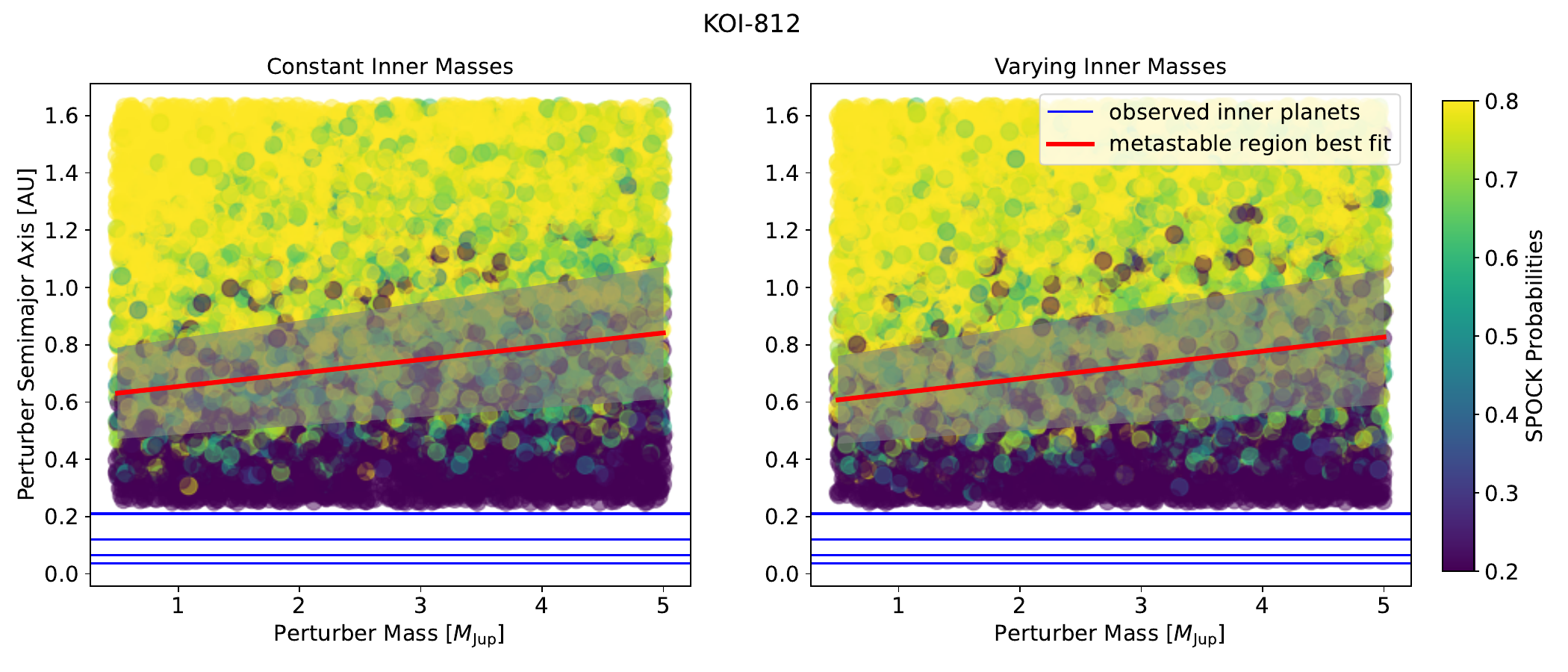}
    
    \centering
    \includegraphics[width=0.8\textwidth]{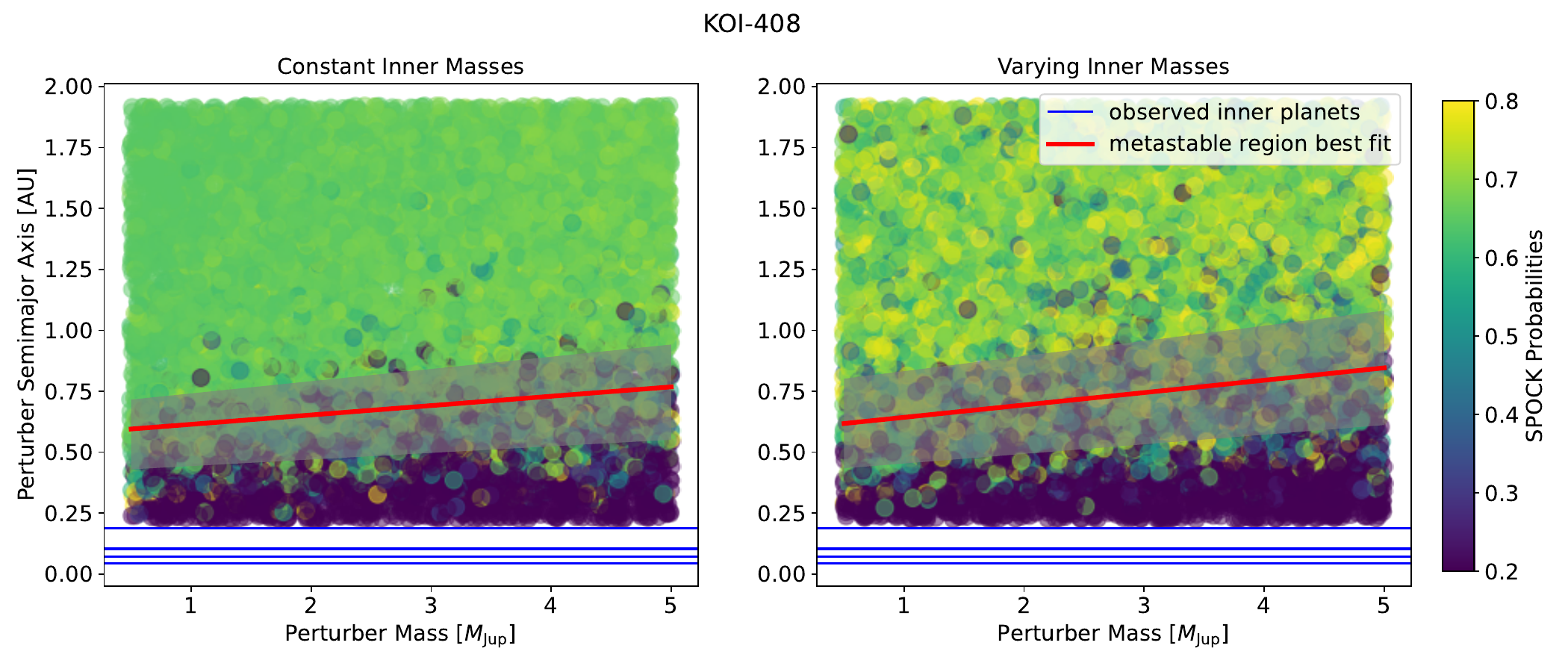}

    \centering
    \includegraphics[width=0.8\textwidth]{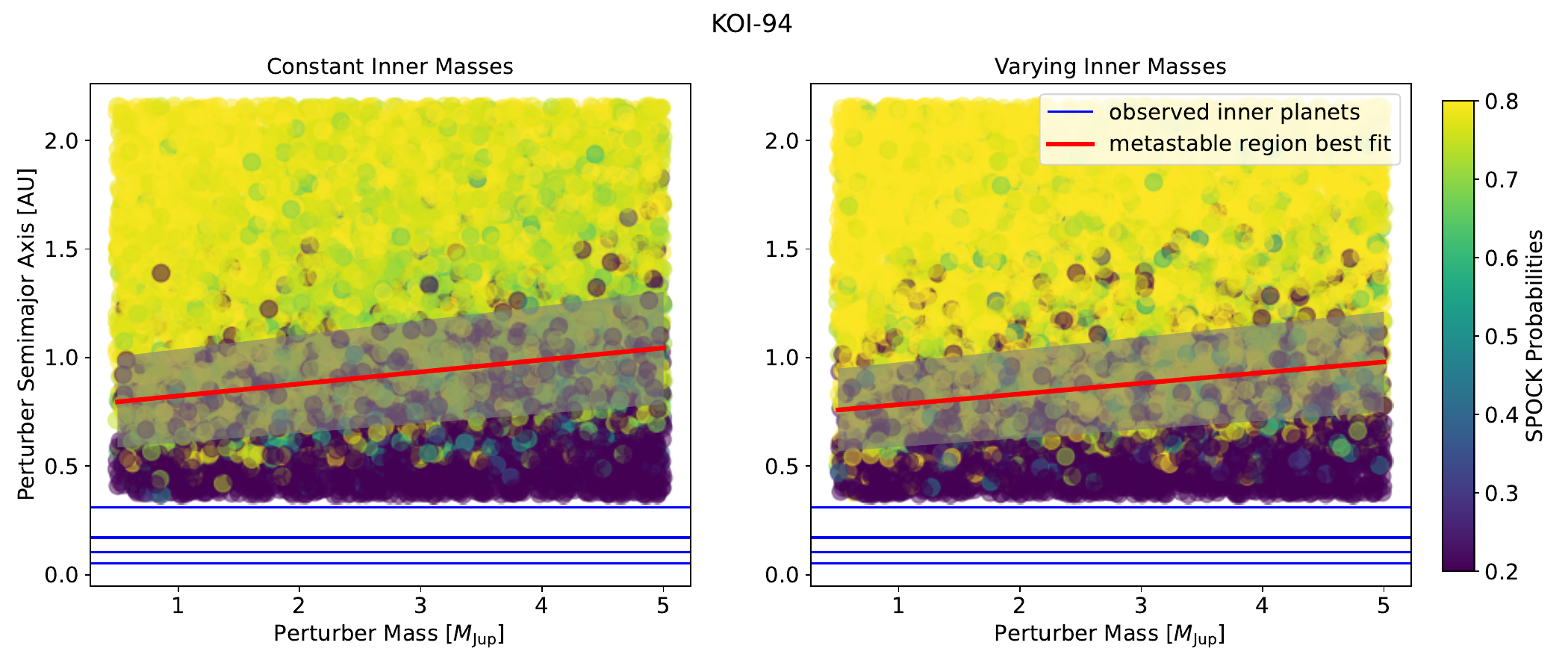}
    \caption{\textbf{Dynamical stability calculations: effect of mass sampling variations.} The scatterplots depict the SPOCK probabilities of stability for three example systems (KOI-812, KOI-408, and KOI-94) with the addition of an outer perturbing planet. Each point corresponds to a single stability calculation with an associated mass and semi-major axis for the outer perturber. The left column illustrates stability calculations in which the inner planet masses are fixed, while the right column has the inner planet masses randomly sampled from a normal distribution as described in Section \ref{sec: dynamical stability calculations w/o perturbers}. The perturber's orbital inclination is randomly sampled from a Rayleigh distribution (Case 1 in Table \ref{tab: parameter space scope}). The colorbar indicates the SPOCK probability of stability. Blue horizontal lines are plotted to indicate the semi-major axes of existing planets. The red line bordered by a translucent gray belt defines the zone corresponding to ``metastability'', the transition from highly unstable to highly stable.}
    \label{fig: KOIs constant vs. varied inner masses}
\end{figure*}

We use the set-up of the inner systems as described in Section \ref{sec: dynamical stability calculations w/o perturbers} and use SPOCK to assess the stability of the augmented systems, consisting of the observed inner planets and the hypothetical outer perturbers. 
For each system, we generate $10,000$ unique parameter combinations and calculate an associated probability of stability for each. Note that the parameters of both the outer perturber and inner planets are randomly sampled, as summarized in Table \ref{tab: parameter space scope}. 

We can visualize the general trends resulting from varying the perturber's physical characteristics by constructing scatterplots of the stability variations as a function of the perturber's parameters. We first create scatterplots of $a_p$ vs. $M_{p,p}$, with $p_{\mathrm{stab}}$ indicated with the colors of the points. Examples are shown for the KOI-812, KOI-408, and KOI-94 systems in Figure \ref{fig: KOIs constant vs. varied inner masses}, which we will discuss at greater length in Section \ref{sec: results}. There is a noticeable division in the $a_p$ vs. $M_{p,p}$ parameter space, which separates regions of low and high probabilities of stability. In all cases, the high probabilities max out at values near the probabilities of stability from the simulations without the perturbers. However, the stable and unstable regions are not starkly divided; we notice an upward sloping transition zone with $p_{\mathrm{stab}} \sim 0.4-0.6$, a region that we will refer to as ``metastable''. This observation motivates the next section's analysis, which is to calculate exactly where this metastable region exists for each system.

\subsection{Metastable transition region}
\label{sec: metastable transition region}

We now describe how we utilize the collection of stability calculations to extract the set of perturber parameters defining the metastable region. We approximate the metastable region as a banded line that divides the stable and unstable regions, and we identify its location using the following numerical procedure. First, we divide the $a_p$ vs. $M_{p,p}$ parameter space into a grid with 10 evenly-spaced segments in $M_{p,p}$ and 40 evenly-spaced segments in $a_p$, and we calculate the mean probability of stability, $\overline{p}_{\mathrm{stab}}$, in each grid cell. The number of grid cells was chosen through trial and error; the spacing ensures that each grid cell is large enough to include many stability values within it but small enough to not smear out the relevant features. The results are not strongly sensitive to these choices.

Next, we use the grid of $\overline{p}_{\mathrm{stab}}$ to find the best-fit line of metastability. We do this by first iterating through columns in the grid (each of which have a fixed $M_{p,p}$) and estimating the value of $a_p$ that lies closest to the center of the metastable region. The metastable region is observed to be a gradual transition between low and high $\overline{p}_{\mathrm{stab}}$, and the curve of $\overline{p}_{\mathrm{stab}}$ vs. $a_p$ for a fixed $M_{p,p}$ (i.e. a single column of the grid) is well-fit by a piecewise linear function of the following form:
\begin{equation}
p(a_p) = 
\begin{cases} 
      \overline{p}_{\mathrm{stab},l} & a_p\leq a_{p,l} \\
      m_{\mathrm{trans}}(a_p - a_{p-l}) + \overline{p}_{\mathrm{stab},l} & a_{p,l} < a_p < a_{p,h} \\
      \overline{p}_{\mathrm{stab},h} & a_p\geq a_{p,h}.
\end{cases}
\end{equation}
Here, $a_{p,l}$ and $a_{p,h}$ are parameters of the function defining the low and high boundaries of the metastable region,  $\overline{p}_{\mathrm{stab},l}$ and $\overline{p}_{\mathrm{stab},h}$ are parameters defining the low and high probabilities at which the curve levels out, and $m_{\mathrm{trans}}=(\overline{p}_{\mathrm{stab},h} - \overline{p}_{\mathrm{stab},l})/(a_{p,h} - a_{p,l})$ is the slope within the metastable region.

We fit the piecewise linear function to the curve of $\overline{p}_{\mathrm{stab}}$ vs. $a_p$ for each column of fixed $M_{p,p}$. We then find the value of $a_p$ for which $p(a_p) = 0.5$. This value, which we denote $a_{p}(p=0.5)$, defines the center of the metastable region within the column. In principle, the edges of the metastable region are defined by $a_{p,l}$ and $a_{p,h}$. However, we find that $a_{p,l}$ is very near to $\min(a_p)$ in most cases, so we define the lower edge of the metastable region to be $a_{p}(p=0.3)$. We define the upper edge of the metastable region to be $a_{p}(p=0.7)$. By repeating this process for each column, we obtain a set of ($M_{p,p}, a_p$) values defining three lines: the center, lower edge, and upper edge of the metastable region. Finally, we perform linear regressions on each curve so as to obtain a smooth demarcation of the metastable region. The centers and edges of the metastable regions are indicated with red lines and gray bands for the examples in Figure \ref{fig: KOIs constant vs. varied inner masses}. 


\section{Results}
\label{sec: results}

We now present the results of our dynamical stability calculations. The first thing we explore is the impact of different sampling choices (Section \ref{sec: sampling variations}). Many parameters are sampled in the exploration of these systems, so it is important to verify that these choices do not strongly impact the results. We will then examine the averaged results for all systems (Section \ref{sec: combined results}). 

\subsection{Effect of sampling variations}
\label{sec: sampling variations}

\begin{figure*}
    \centering
    \includegraphics[width=0.8\textwidth]{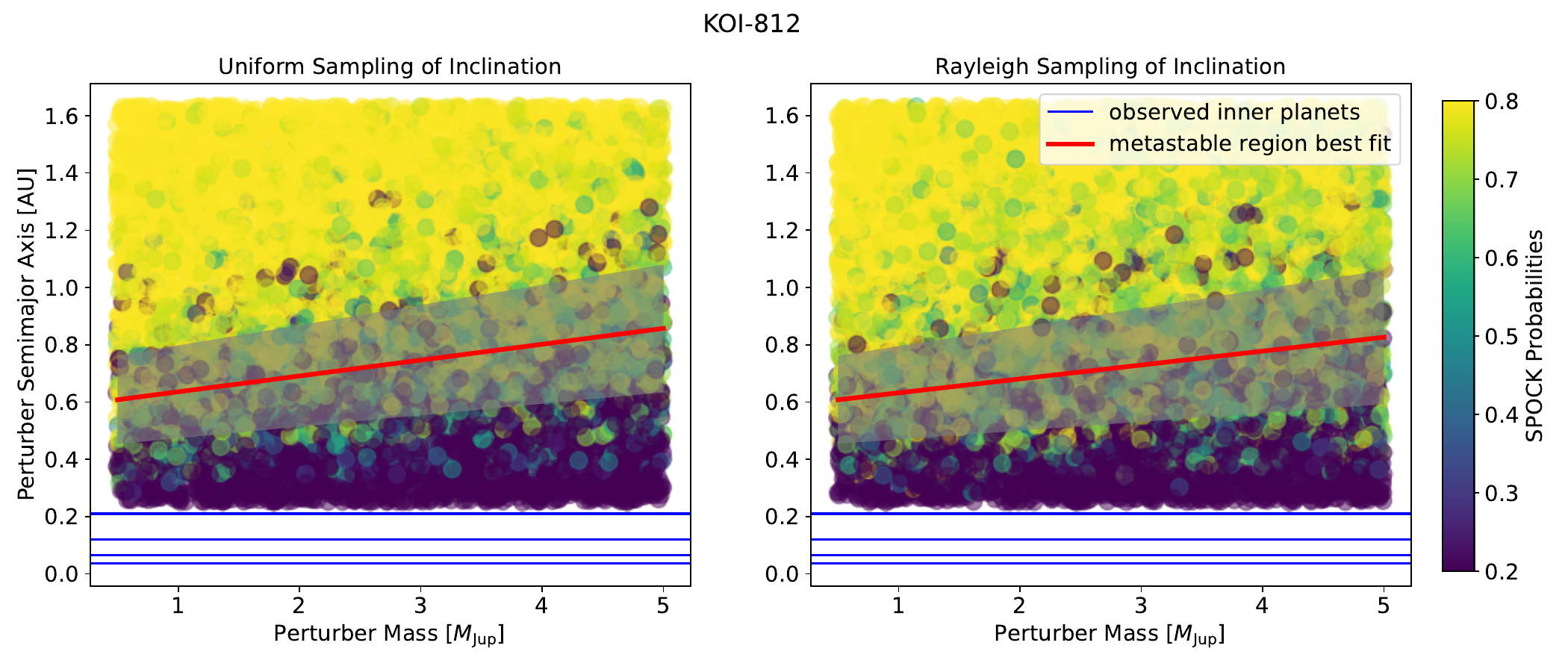}

    \centering
    \includegraphics[width=0.8\textwidth]{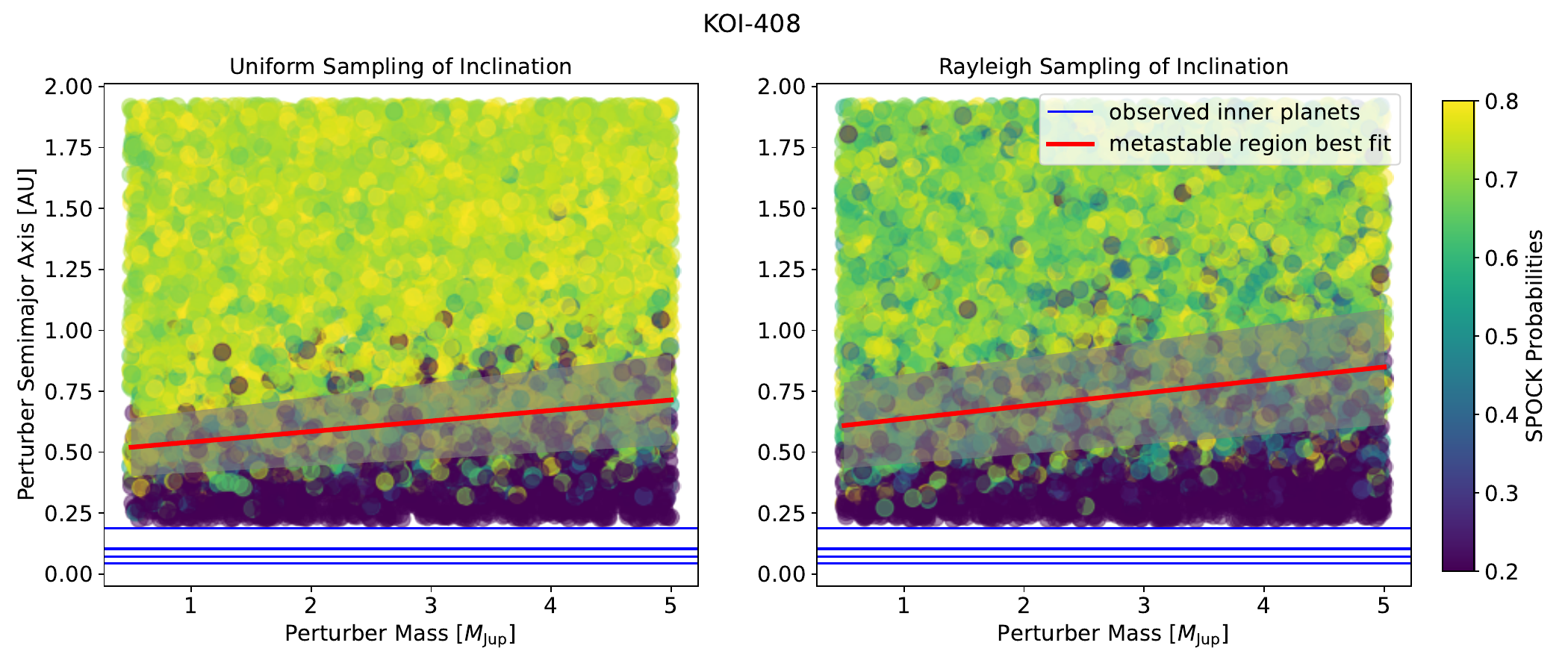}

    \centering
    \includegraphics[width=0.8\textwidth]{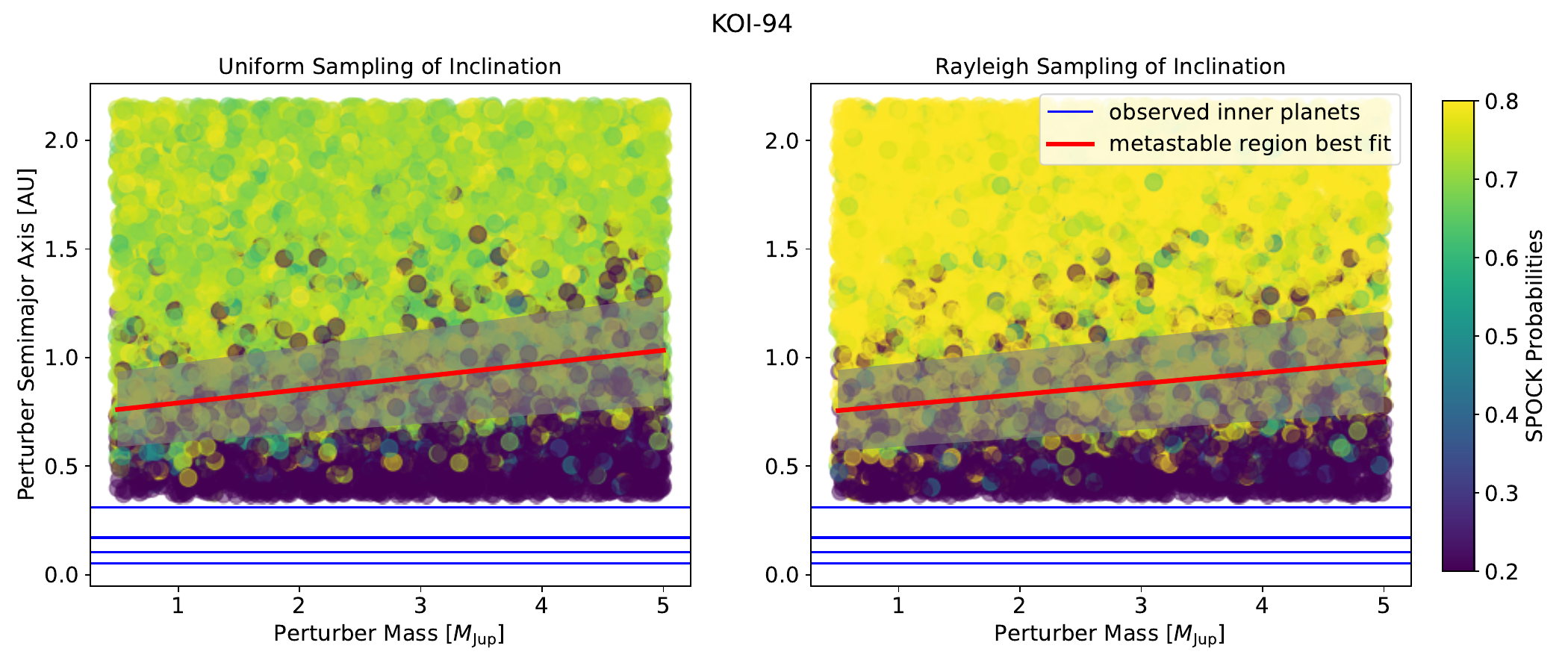}
    \caption{\textbf{Dynamical stability calculations: effect of inclination sampling variations.} The plotting scheme and example systems are identical to Figure \ref{fig: KOIs constant vs. varied inner masses}, except we now explore different methods of sampling the outer perturber's inclination. The left column indicate uniform sampling $i_p \sim \mathrm{Uniform}[0^{\circ}, 40^{\circ}]$, and the right indicates Rayleigh-distributed sampling ${i_p \sim \mathrm{Rayleigh}(10^{\circ})}$. In all cases, the inner planet masses are randomly sampled from a normal distribution as in the right column of Figure \ref{fig: KOIs constant vs. varied inner masses}.}
    \label{fig: KOIs uniform vs. rayleigh inclinations}
\end{figure*}

\begin{figure*}
    \centering
    \includegraphics[width=0.8\textwidth]{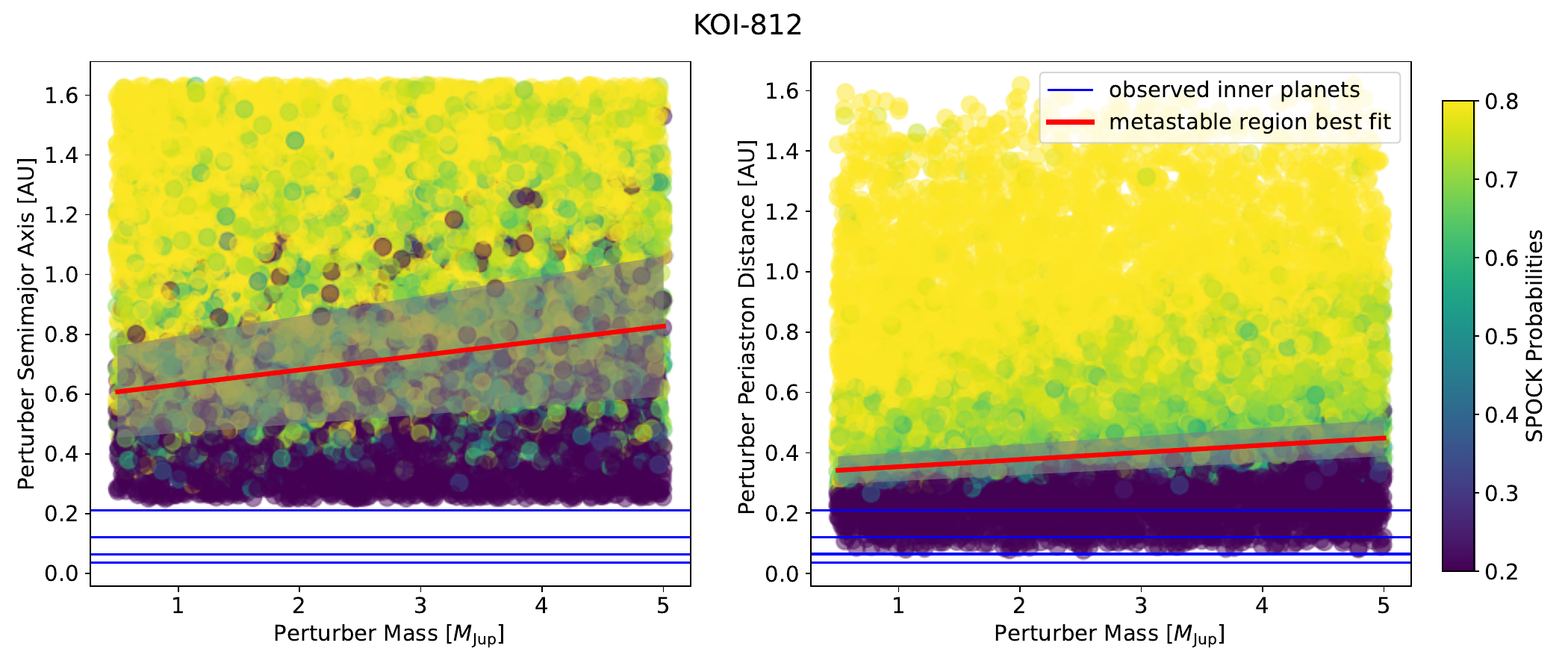}

    \centering
    \includegraphics[width=0.8\textwidth]{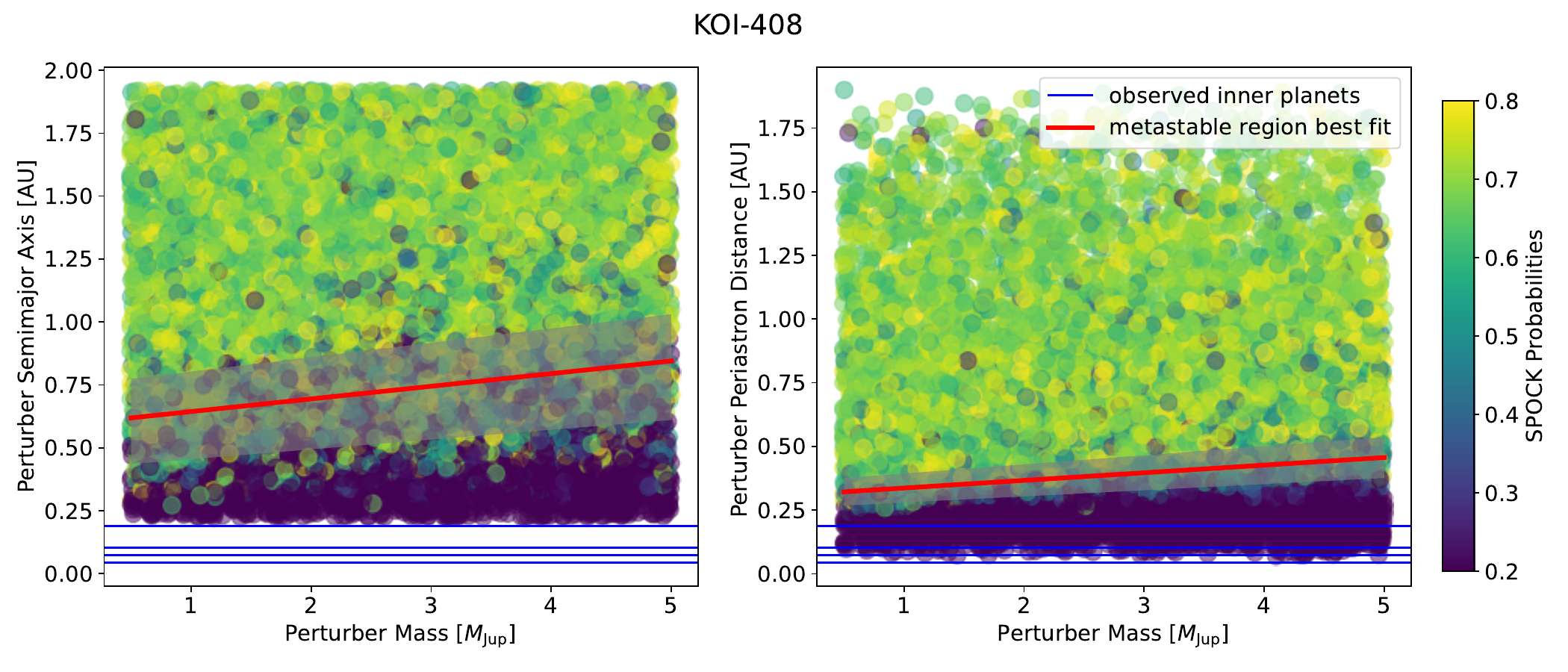}

    \centering
    \includegraphics[width=0.8\textwidth]{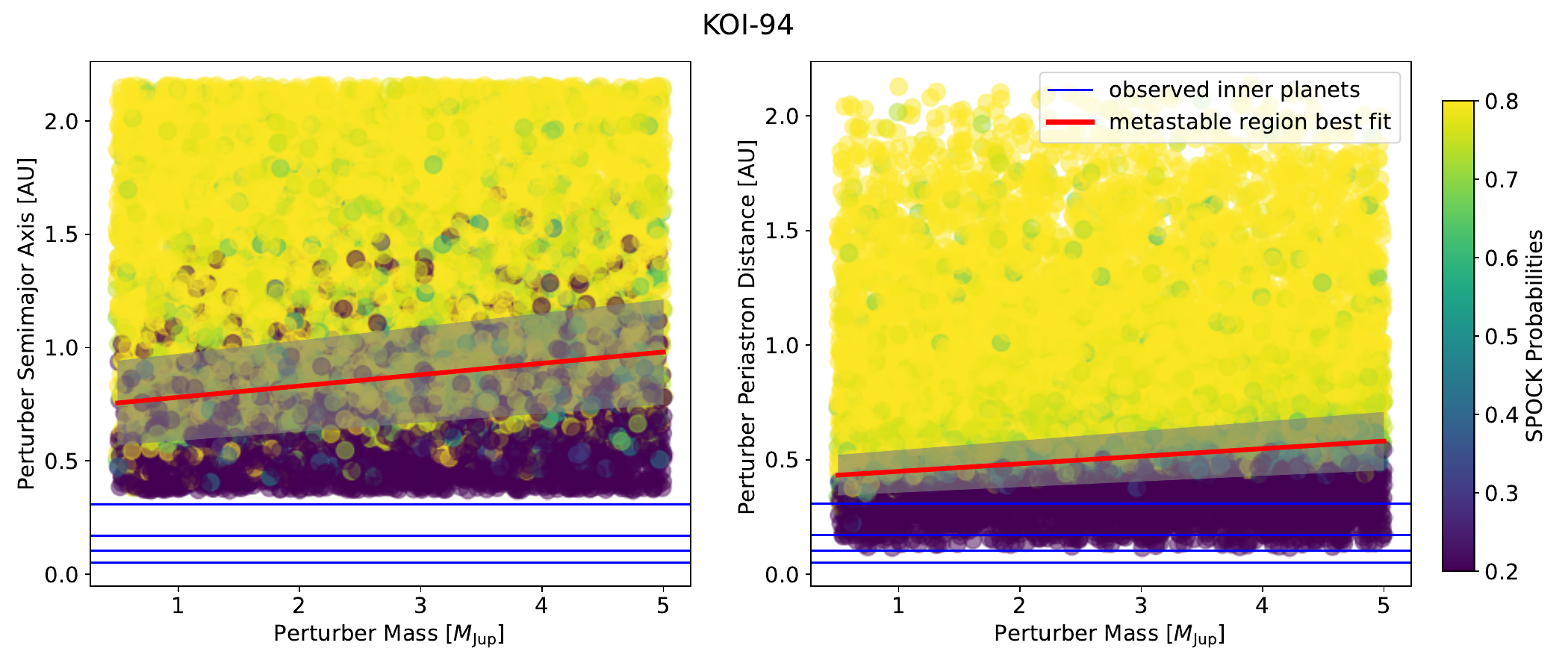}
    \caption{\textbf{Dynamical stability calculations: dependence on perturber semi-major axis vs. periastron distance.} The plotting scheme and example systems are identical to Figure \ref{fig: KOIs constant vs. varied inner masses} and Figure \ref{fig: KOIs uniform vs. rayleigh inclinations}, but now the left column shows the semi-major axis vs. mass, while the right column shows the perturber periastron distance vs. mass. In all cases, the inner planet masses are randomly sampled from a normal distribution as in the right column of Figure \ref{fig: KOIs constant vs. varied inner masses} and the orbital inclination of the outer perturber is Rayleigh-distributed.}
    \label{fig: KOIs semi-major axis vs. periastron distance}
\end{figure*}

Figure \ref{fig: KOIs constant vs. varied inner masses} shows the impact of sampling the inner planet masses, where the left and right columns show the results when the masses are held fixed and varied, respectively. Sampling the inner planet masses makes some systems a bit more stable and other systems a bit more unstable. However, for most systems, there is very little overall change in the location and width of the metastable region, suggesting that the results are not strongly sensitive to the inner planet mass sampling scheme. 

Figure \ref{fig: KOIs uniform vs. rayleigh inclinations} shows the impact of the different methods of sampling the outer perturber's inclination. The left and right columns show the results of the uniformly-distributed and Rayleigh-distributed sampling (recall the two cases in Table \ref{tab: parameter space scope}). As in Figure \ref{fig: KOIs constant vs. varied inner masses}, there are some slight differences between the two versions, but overall the general features and the locations and widths of the metastable regions are the same regardless of the sampling method.

Finally, Figure \ref{fig: KOIs semi-major axis vs. periastron distance} displays the results for the same three example systems in a different parameter space. The left column is the same as the right column of Figure \ref{fig: KOIs uniform vs. rayleigh inclinations}, displaying perturber semi-major axis vs. mass. (Here we use Rayleigh-distributed perturber inclinations.) The right column replaces the y-axis with the perturber periastron distance, $a_p (1-e_p)$. It is immediately clear that plotting the results in this space yields a sharper transition from the unstable to stable region and thus a smaller metastable region. This makes physical sense because the inner planets are more sensitive to the perturber's close approach distance than its semi-major axis.

\subsection{Combined constraints}
\label{sec: combined results}

\begin{figure}
    \centering
    \includegraphics[width=0.95\columnwidth]{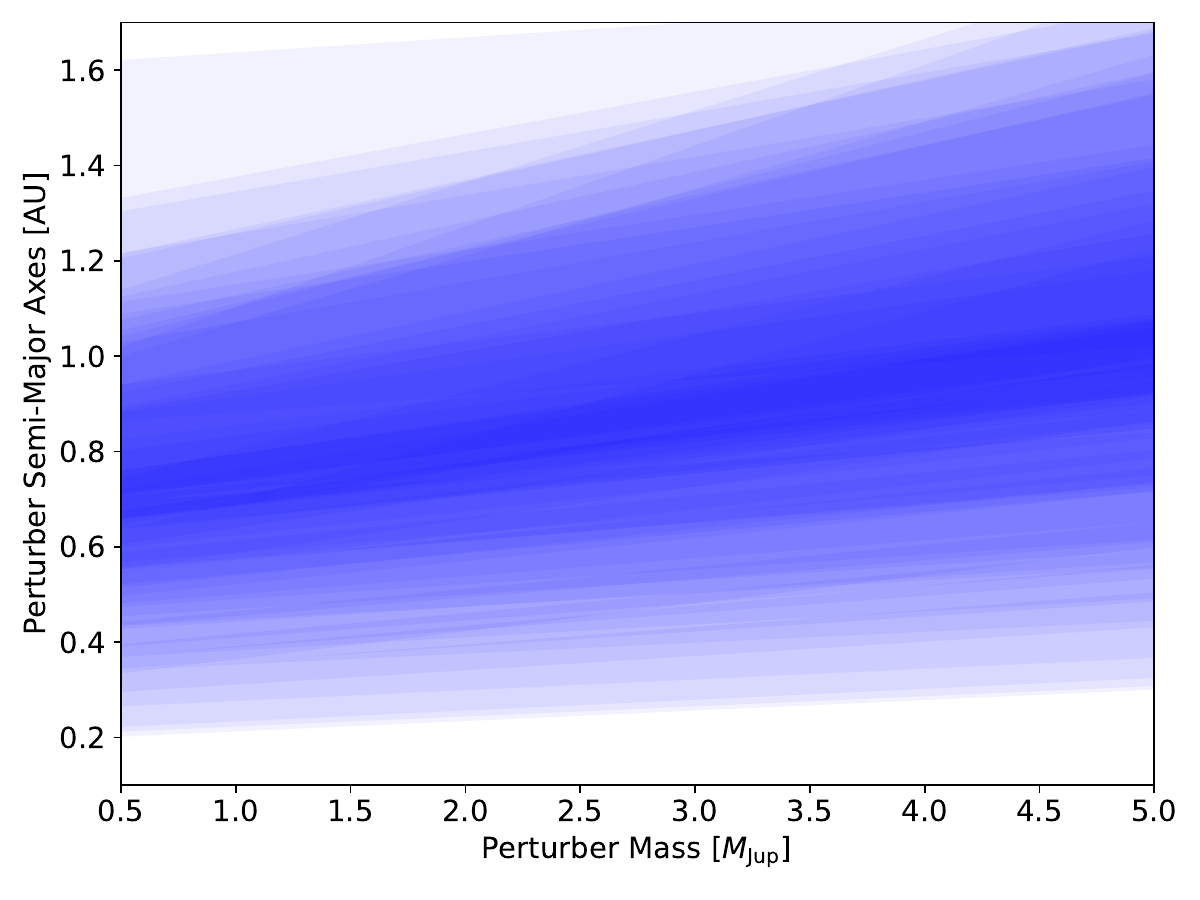}
    \includegraphics[width=0.95\columnwidth]{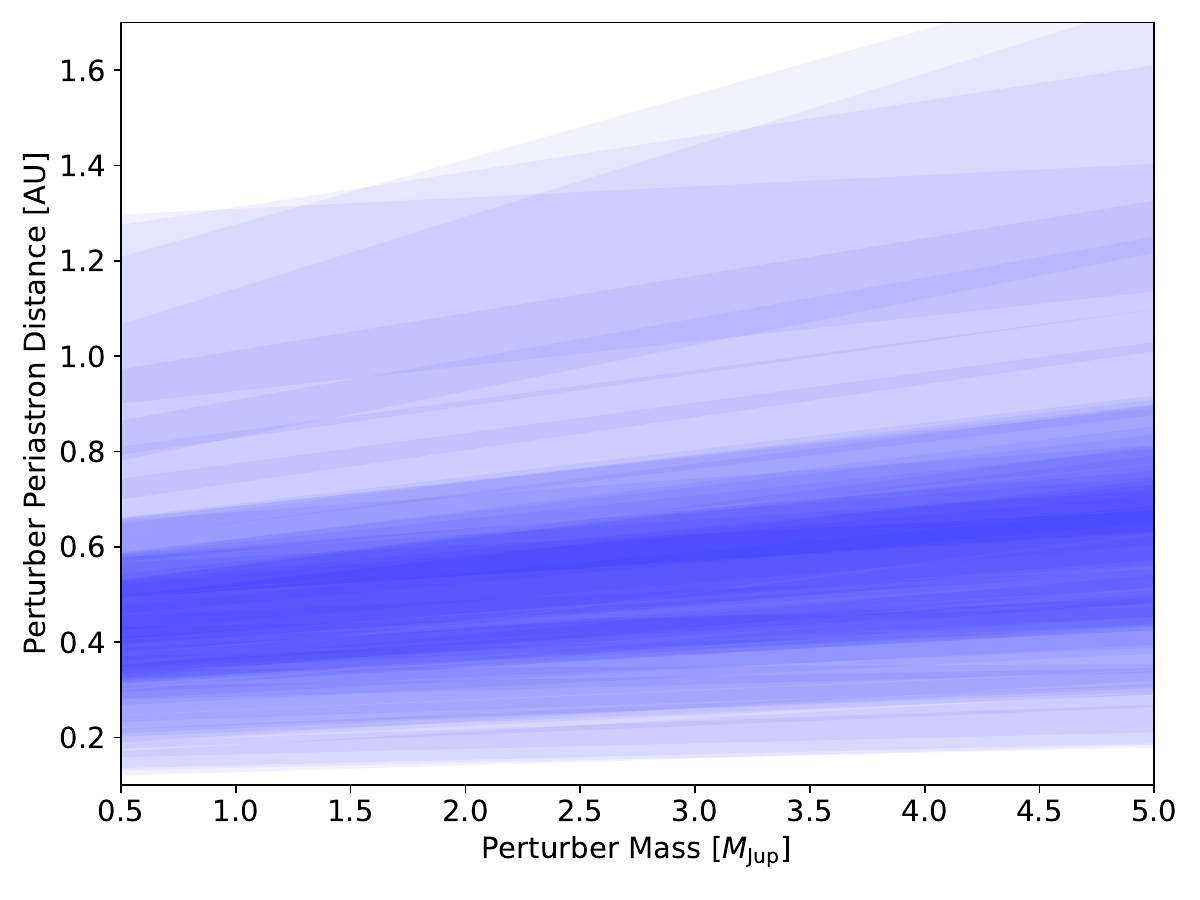}
    \includegraphics[width=0.95\columnwidth]{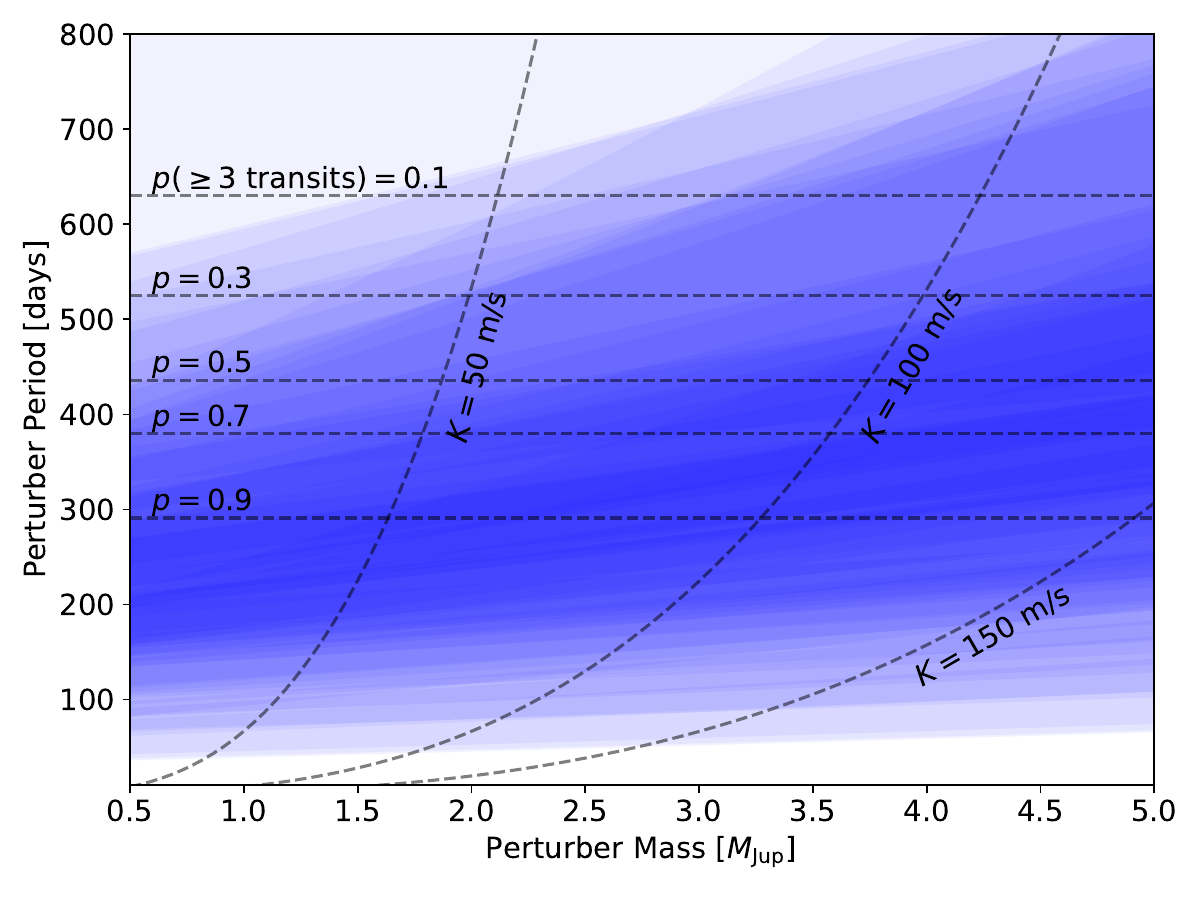}
    \caption{\textbf{Aggregate results of stability calculations.} We display the ``metastable regions'' (grey belts in Figures \ref{fig: KOIs constant vs. varied inner masses}-\ref{fig: KOIs semi-major axis vs. periastron distance}) for all systems in our sample. The top, middle, and bottom panels show the results in perturber semi-major axis, periastron distance, and period vs. mass. The depicted metastable regions are taken from the results in which the inner planet masses are randomly sampled from a normal distribution around the mean and the orbital inclination of the outer perturber is Rayleigh-distributed. The darkest blue indicates where there is greatest overlap among the various systems. In the bottom plot, we include contours of the outer perturber's RV semi-amplitude (curved lines) and, if it was transiting, the probability that it would transit three or more times in Kepler data (horizontal lines). }
    \label{fig: semi-major axes of all metastable}
\end{figure}

We now combine the results for all systems in order to explore the influence of distant perturbing planets on the ensemble of compact multi-planet systems. Figure \ref{fig: semi-major axes of all metastable} shows the extracted metastable regions for all systems. The darkest color indicates where there is the greatest overlap among the various systems. The metastable region is concentrated in the period range $P \sim 100 - 500$ days ($a \sim 0.4 - 1.2$ AU for $M_{\star} = M_{\odot}$). 

We also display information on the detectability of the perturbing planet using RV and Kepler transit data. As for RV detection, we plot contours of constant RV semi-amplitude assuming $M_{\star} = M_{\odot}$ and $e=0$. For the vast majority of the parameter space, the perturber would be easily detected in RV data of a sufficiently long baseline, since the semi-amplitude is much greater than typical RV uncertainties.

As for transit detection, we calculate the conditional probability that, provided the perturber was transiting, it would transit its host star at least three times in Kepler data. We obtain this estimate using a data product provided by Kepler DR25 \citep{2018ApJS..235...38T, koidr25} called the window function, which is the fraction of unique transit ephemeris epochs that permit three or more transits to be observed as a function of orbital period \citep{2017ksci.rept...14B}. This quantity is complex due to the data gaps within the limited span of Kepler photometry. Kepler DR25 provided tabulations of window functions for each star as a function of orbital period and transit duration. We use the downsampled tables provided by \cite{2019AJ....158..109H} to extract the window function data for each of the host stars in our sample and obtain the values for periods in the range $P = 1-800$ days, picking the transit duration such that it aligns with expectations of circular orbits. We then average over the different stars in our sample. 

Using the averaged window function data, we calculate the orbital periods at which the window function is equal to 0.1, 0.3, 0.5, 0.7, and 0.9. The periods are indicated with the horizontal lines in Figure \ref{fig: semi-major axes of all metastable}. Since the perturbing planets would have large radii (and thus large transit signal-to-noise ratios), the main factor in their detection by the Kepler pipeline would simply be whether they have three or more transits. Accordingly, the window function effectively provides the probability that the perturbing planets would be detected if they were on transiting orbits. We observe that roughly half of the combined metastable region would have $p(\geq 3 \ \mathrm{transits}) = 0.7$, indicating that many of them would be detected in Kepler data if they existed and were transiting. 


\section{Discussion}
\label{sec: discussion}

\subsection{Implications for the edge-of-the-multis}
\label{sec:Implications for the edge-of-the-multis}
The results of Figure \ref{fig: semi-major axes of all metastable} indicate that perturbing planets with parameters falling in the average metastable region would be readily detectable with RV data (of a sufficiently long baseline) and transit data. What are the implications of this? We argue that it indicates that \textit{distant giant planets likely cannot explain the edge-of-the-multis}, according to the following logic. First, the metastable region delineates two regimes: (1) the unstable regime in which the distant giant planet is strongly coupled to the inner planets, and they cannot maintain dynamical stability, and (2) the stable regime in which the distant giant planet's perturbative influence is weak relative to the coupling among the inner planets, and they maintain stability. Accordingly, the metastable region (or, more accurately, the region just slightly wide of it) is an approximate indicator of the distant giant ``edge-sculpting'' properties, that is, the range of possible properties that a perturbing planet could have if it is responsible for sculpting the outer edge of the inner system. To see this, consider a system where the perturbing planet has a period larger than the metastable location. With such properties, the inner system could potentially have another planet added to its outer edge and still be stable, so the giant planet would not be consistent with sculpting the edge of the observed system.

If the metastable region thus approximates the parameters of edge-sculpting perturbing planets, and if most of the metastable region would be readily detectable with sufficient radial velocity and transit data, then edge-sculpting perturbing planets would generally be detectable. However, such perturbing planets are not widely detected in this region of parameter space. Here we will briefly review the observational constraints. 

As for transits, all but one of the compact multis in our sample are devoid of a transiting giant planet with a period in the range $P=100-500$ days confirmed by the Kepler search pipeline. One exception is the Kepler-90 (KOI-351) system, which contains a sun-like star and eight confirmed transiting planets, the outermost of which has period $P=332$ days and mass $203\pm 5 \ M_{\oplus}$ \citep{2021AJ....161..202L}. This outer planet may be partially sculpting the edge of the inner seven planets. Another possible exception is Kepler-87 (KOI-1574), which contains two planets beyond 100 days, one planet at 115 days with mass $324 \pm 9 \ M_{\oplus}$ and one planet at 191 days with mass $6.4 \pm 0.8 \ M_{\oplus}$ \citep{2014A&A...561A.103O}. The planet at 115 days could be responsible for dynamical perturbations, but it does not strictly fit our definition of edge-sculpting giant planets, since the system contains only two planets interior to it and a small exterior planet.

Beyond detections from the Kepler pipeline, \cite{2016AJ....152..206F} performed a systematic search for transiting long-period giant planets in Kepler data, requiring only one or two transits (rather than the usual $\geq 3$). They found 16 planet candidates, one of which is in a system in our sample. Kepler-154 (KOI-435) contains five inner super-Earths/sub-Neptunes with periods ranging from $3 - 62$ days and an outer transiting giant with $P = 4.3^{+4.7}_{-1.3}$ yr and $R_p = 0.83^{+0.12}_{-0.11} \ R_{\mathrm{Jup}}$ \citep{2016AJ....152..206F}. However, this outer giant is too far from the inner system to be responsible for sculpting the edge. Using injection-recovery experiments, \cite{2016AJ....152..206F} found that their search method was $\sim70\%$ complete to planets with periods $P \lesssim 3$ yr and $R_p \gtrsim 0.5 \ R_{\mathrm{Jup}}$. Roughly speaking, this indicates that they would be $\sim70\%$ complete to any transiting edge-sculpting giant planets orbiting the stars in our sample, of which none were detected other than in Kepler-154. 

The lack of many transit detections is suggestive but inconclusive, since the intrinsic transit probability is very small at long periods, and it depends on the unknown orbital inclinations. However, we can still obtain a rough estimate of the number of edge-sculpting giant planets we would expect to be transiting in our sample. We use our calculated metastable region for each system\footnote{For systems without SPOCK calculations (see Section \ref{sec: dynamical stability calculations w/o perturbers}), we randomly pick a metastable region from the rest of the sample.}
and randomly sample an orbital period for a giant planet within it.  We then sample its inclination as $i \sim \mathrm{Uniform}[80^{\circ},100^{\circ}]$. This assumes that the inner system is approximately edge-on ($i \approx 90^{\circ}$), and it roughly agrees with observed constraints on the mutual inclinations between inner super-Earths and outer cold Jupiters. \cite{2020AJ....159...38M} found that these mutual inclinations tend to be small, especially for inner systems with higher transit multiplicity. Using the sampled period and inclination, we determine whether or not the planet would be transiting. We then sum up the number of transiting planets across the full sample of 64 systems. Finally, we repeat this process 1,000 times to create a distribution of the number of transiting planets. This yields an expectation of $2.0 \pm 1.3$ transiting planets across the 64 systems. The percentages of cases with 0, 1, 2, 3, 4, and 5 transiting planets are 18\%, 32\%, 25\%, 16\%, 7\%, and 2\%. This indicates that it would be fairly likely to observe zero transiting giant planets, even if all of the systems hosted edge-sculpting giant planets. Constraints from transits are thus not very powerful; RVs are likely to be more informative. 

As for RV detections, several constraints have recently become available through long term HARPS-N and Keck/HIRES surveys presented by \cite{2023arXiv230405773B}
and \cite{2023arXiv230400071W}, respectively. As for \cite{2023arXiv230405773B}, their sample of 38 Kepler and K2 systems of small planets has only three systems that overlap with our sample. None of those have new detected planets or long-term trends. The Kepler Giant Planet Survey \citep{2023arXiv230400071W} observed 63 systems, 13 of which are common to our sample. Of those systems, nine have no new detections or long-term trends, and the data places strong constraints on the existence of additional long-period planets. In most cases, the RV coverage places a $3\sigma$ upper limit of $M \sin i \lesssim 0.5 \ M_{\mathrm{Jup}}$ at 5 AU (although there are variations from system to system). The existence of edge-sculpting perturbers can thus be ruled out in these systems. 

The remaining four systems have RV-detected non-transiting companions: Kepler-20 (KOI-70), Kepler-106 (KOI-116), Kepler-1130 (KOI-2169), and Kepler-444 (KOI-3158). In Kepler-20 (KOI-70), there is a detected non-transiting planet with $P = 34.9$ days and $M\sin i = 21.0 \pm 3.4 \ M_{\oplus}$ in between the planets with $P = 19.6$ days and $P = 77.6$ days. This planet was initially discovered by \cite{2016AJ....152..160B}, although it remains possible that the signal is a false positive due to stellar activity \citep{2020AJ....159...23N}. Kepler-106 (KOI-116) has four small transiting planets and reveals peaks in the RV residual periodogram near P = 90, 180, and 365 days. The veracity of these signals is uncertain because they could be driven by the seasonal window function. However, if there is a planet at 90 days, it would have $M \sin i = 46 \pm 6 \ M_{\oplus}$. Kepler-1130 (KOI-2169) contains a stellar-mass companion with $P = 14000 \pm 1800$ days ($\sim12$ AU), $M \sin i = 218 \pm 3 \ M_{\mathrm{Jup}}$, and $e = 0.65 \pm 0.024$. The periapse distance is $\sim4$ AU. Finally, Kepler-444 (KOI-3158) also contains an M-dwarf binary companion (components B and C) first characterized by \cite{2016ApJ...817...80D}. The orbital solution in \cite{2023arXiv230400071W} is consistent with a recent analysis by \cite{2023AJ....165...73Z} and finds $P = 87000 \pm 3000$ days ($\sim 52$ AU), $M \sin i = 629 \pm 21 \ M_{\mathrm{Jup}}$, and $e = 0.55 \pm 0.05$. The periapse distance is $\sim 23$ AU. 

Among the systems just described, the Kepler-1130 system contains the only perturber with properties that might be consistent with sculpting the outer edge of the inner system, although the perturber is a star, not a giant planet as considered in this work. It is worth noting that the two systems with stellar-mass companions, Kepler-1130 (KOI-2169) and Kepler-444 (KOI-3158), contain some of the smallest inner planets in our sample. The planets all have smaller radii than Earth, and \cite{2017ApJ...838L..11M} constrained the masses of the Kepler-444 planets to be similar to that of Mars. It is thought that the stellar companions limited the extent of the protoplanetary disk in which the inner planets formed, thus limiting the solid material for planet formation \citep[e.g.][]{2023AJ....165...73Z}. However, this is a different process than the dynamical sculpting we have been considering in this paper. 

Altogether, the existing constraints from both transits and radial velocities suggest a lack of perturbing planets that are close enough to the inner compact multis to sculpt their outer edges. The observational constraints are incomplete, but they are sufficient to conclude that distant perturbers likely contribute very little to the observed edge-of-the-multis. It is important to clarify that the preceding discussion is not intended to be exhaustive, as it neglected several complications. For instance, we neglected the possibility that a perturbing planet that is not close enough to an inner system to render it dynamically unstable could still potentially increase the mutual inclinations among that inner system and reduce the transit multiplicity. This would be a different form of edge-sculpting.  Moreover, our estimation of the expected number of transiting edge-sculpting giant planets used several simplifying assumptions.  It is beyond the scope of this work to perform a thorough estimate of how many giant planets should have been detected in existing transit and radial velocity observations if they are the primary cause for the edge-of-the-multis. We leave a more thorough accounting for these detection statistics to future work.

\subsection{Alternative theories for the edge-of-the-multis}
If distant giant planets are not the explanation of the edge-of-the-multis, we must turn towards alternative hypotheses. Several theories were outlined in \cite{2022AJ....164...72M}, and additional theories have since been proposed. We briefly recap them here. First, the formation process of compact multis may naturally confine a system of $\sim4-6$ inner planets to the $\lesssim 0.5-1$ AU region, with a truncation beyond them. For instance, the theory of ``inside-out planet formation'' \citep{2014ApJ...780...53C, 2015ApJ...798L..32C} posits that planets may form sequentially from successive gravitational instabilities of rings of pebbles. The rings form via pebble drift due to gas drag and build up at the pressure maximum associated with the dead-zone boundary. After the planets form from the pebble ring, they migrate inwards, the dead-zone boundary and pressure maximum retreat, and the process repeats. A similar theory of successive formation of super-Earths from narrow rings of solid material was recently proposed by \cite{2023NatAs...7..330B}. In this framework, planetesimals form rapidly at the silicate sublimation line and grow through pairwise collisions, until they achieve terminal proto-planet masses regulated by isolation and migrate inwards. Systems that host distant giant planets have also been shown to potentially drive the formation of high-density rings of planetesimals due to the sweeping of secular resonances that transport planetesimals into the inner disk \citep{2023arXiv230402045B}. All of these theories naturally predict an outer edge of compact multis near the observed edge. 

Even if the planets in compact multis do not form at well-defined discrete locations in the inner disk, an edge-of-the-multis is also predicted by theories of formation at larger orbital distances ($\gtrsim 1$ AU) followed by inward migration. Specifically, \cite{2022ApJ...937...53Z} showed that migration traps (generated by regions where the outward corotation torque dominates over the inward Lindblad torque) yield planetary systems that are bifurcated into two groups at small and larger periods with a gap at $\sim100-300$ days. 

To summarize, these theories predict that the formation of compact multis is a balance between accretion and orbital migration that results in a compact system of planets confined to the inner $\sim0.5-1$ AU, with a truncation just outside. We reviewed these hypotheses here for the sake of promoting additional and probable arguments based on our findings, but further investigation is beyond the scope of this paper. Future work on both planet formation modeling and observational characterization of the $\gtrsim 1$ AU regions of these systems should help determine which, if any, of these mechanisms are dominant in sculpting the edge-of-the-multis.

\section{Conclusion}
\label{sec: conclusion}

This work was motivated by the recent finding that the outer edges of compact multi-planet systems appear to truncate at smaller periods than expected from geometric and detection biases alone \citep{2022AJ....164...72M}. The ``edge-of-the-multis'' suggests the existence of some truncation (i.e. occurrence rate fall-off) or transition (i.e. to smaller and/or more widely-spaced planets) in the outer regions ($P \gtrsim 100$ days) of compact multis. Here we investigated the question of whether the edge-of-the-multis could be caused by the dynamical influence of distant giant planets, which are thought to be common in systems with inner super-Earths/sub-Neptunes \citep{2018AJ....156...92Z, 2019AJ....157...52B}. 

We considered a sample of Kepler compact multi-planet systems with four or more observed planets (Figure \ref{fig: planet sample}). We explored the dynamical stability of the observed systems in the presence of hypothetical exterior giant planets. The SPOCK machine learning stability classifier \citep{2020PNAS..11718194T} was used to compute the probability of dynamical stability of a given orbital configuration. We randomly sampled the perturber period and mass in the range $P_p \lesssim 1000$ days and $M_{p,p} \sim 0.5 - 5 \ M_{\mathrm{Jup}}$. We tested multiple sampling schemes for other parameters (e.g. the perturber inclination) and found minimal sensitivity to these choices (Figures \ref{fig: KOIs constant vs. varied inner masses} and \ref{fig: KOIs uniform vs. rayleigh inclinations}). 

Given a set of perturber configurations for each system, we identified the ``metastable region'', the region in perturber mass/semi-major axis space that divides the stable and unstable regimes (e.g. Figure \ref{fig: KOIs constant vs. varied inner masses}). The metastable region associated with an observed inner system approximately defines the parameter regime in which a perturber would dynamically sculpt the edge of the system. That is, a perturber with a wider orbit than the metastable region could leave room for additional (undetected) inner planets and thus would not be an ``edge-sculpting'' perturber. We found the metastable region to be in the range $P\sim100-500$ days, with the strongest concentration around $P\sim200$ days for $M_{p,p} \sim 0.5 \ M_{\mathrm{Jup}}$ and $P\sim400$ days for $M_{p,p} = 5 \ M_{\mathrm{Jup}}$.

We explored the detectability of perturbing planets with ``edge-sculpting'' properties, finding that they would generally be fairly easy to detect. If such perturbers were transiting in Kepler data, many of them would transit at least three times. Roughly half would have a $>0.7$ probability of three or more transits. However, transit probabilities are small for long-period planets, and they depend on the unknown orbital inclinations. A thorough model of predicted transit yields would be necessary for a complete characterization of the transit detectability. As for RV detection, the vast majority of parameter space of edge-sculpting perturbers would have RV semi-amplitudes well above 50 m/s.

Despite their apparent detectability, ``edge-sculpting'' perturbing planets have not been identified in many compact multi systems. We reviewed the current observational constraints, finding one system (Kepler-90) with a transiting outer giant planet that may be consistent with edge sculpting. As for radial velocity constraints, about 14 (perhaps more) of the 64 systems in our sample have sufficient RV coverage to rule out perturbing planets with edge-sculpting properties. The observational evidence is incomplete, but it clearly does not point towards a large number of edge-sculpting perturbers. Future observations will strengthen this result. 

Taken as a whole, our results indicate that distant giant planets are unlikely to play a significant role in sculpting the outer edges of compact multis. Our finding narrows the search for the true cause of this phenomenon. The most likely hypothesis is that the edge-of-the-multis is a signature of the formation process, which may be relatively insensitive to the presence or absence of distant perturbing planets. Future theoretical and observational efforts may test this interpretation. In particular, we expect the upcoming PLATO mission \citep{2014ExA....38..249R} will yield further constraints on the outer architectures of compact multis, especially if it revisits the Kepler field for a long baseline.

\section{Acknowledgements}
We thank Dan Tamayo and the anonymous reviewer for their helpful comments and suggestions. We thank the MIT Undergraduate Research Opportunities Program (UROP) for their support in making this research possible. This
research has made use of the NASA Exoplanet Archive, which is operated by the California Institute of Technology, under contract with the National Aeronautics and Space Administration under the Exoplanet Exploration Program. Some of the calculations presented in this paper were performed on the MIT Engaging cluster at the Massachusetts Green High Performance Computing Center (MGHPCC) facility. We gratefully MGHPCC for their computational resources and support. 

\newpage
\appendix

\section{Testing the validity of the SPOCK calculations}
\label{sec: testing SPOCK}

Here we test the validity of the SPOCK calculations for our systems of inner super-Earths and outer giant planets by verifying that the metastable region agrees with the transition between stable and unstable regions identified by alternative metrics. We first use the Mean Exponential Growth factor of Nearby Orbits (MEGNO) chaos indicator, which quantifies the degree of divergence of initially closely-separated trajectories in phase space \citep{2000A&AS..147..205C,
2003PhyD..182..151C}. The time-averaged MEGNO value distinguishes between stable/regular and unstable/chaotic orbital motion and thus provides an independent calculation of the stability regions. We considered one of our systems (KOI-812) as a case study and generated a similar $a_p$ vs. $M_{p,p}$ map with 1,000 unique parameter combinations. Overall, we find good agreement between the SPOCK and MEGNO maps, which increases confidence in the validity of our SPOCK calculations.

We also check our results using an analytic stability criterion. Various criteria have been developed for compact multis with distant perturbers \citep[e.g.][]{2018MNRAS.478..197P, 2019MNRAS.482.4146D, 2021AJ....162..220T}. We use the results from \cite{2018MNRAS.478..197P} and compute the averaged coupling parameter, $\bar{\epsilon}$ (their equation 60), which quantifies the extent to which the inner planets are coupled to the giant planet ($\bar{\epsilon} \gg 1$) versus each other ($\bar{\epsilon} \ll 1$). The coupling parameter is an approximate stability criterion, since the degree of eccentricity and inclination excitation in the inner system is much higher when they are tightly-coupled to the giant planet. We compute $\bar{\epsilon}$ for the KOI-812 system and find that the $\bar{\epsilon}=1$ transition line agrees well with the SPOCK results. Given these numerical and analytic checks, we are confident in using SPOCK for our stability calculations.  

\bibliographystyle{aasjournal}
\bibliography{main}

\end{document}